\documentclass[12pt]{article}

\usepackage{amssymb}
\usepackage{amsmath}
\usepackage{amscd}
\usepackage{latexsym}

\usepackage{cite}

\usepackage{graphicx}
\usepackage{array}

\newcommand{\be}{\begin{equation}}
\newcommand{\ee}{\end{equation}}

\newcommand{\dlt}{\delta}

\newcommand{\cH}{{\cal H}}
\newcommand{\br}{{\bf r}}

\newcommand{\ra}{\rightarrow}

\newcommand{\bt}{\beta}
\newcommand{\vp}{\varphi}

\newcommand{\al}{\alpha}

\newcommand{\gm}{\gamma}
\newcommand{\om}{\omega}
\newcommand{\Om}{\Omega}

\newcommand{\dgr}{\dagger}

\newcommand{\rgl}{\rangle}
\newcommand{\lgl}{\langle}
\newcommand{\prt}{\partial}

\begin{document}

\begin{center}

{\Large{\bf Spontaneous symmetry breaking under Bose-Einstein condensation} \\ [5mm]

V.I. Yukalov$^{1,2}$ } \\ [3mm]

{\it $^1$Bogolubov Laboratory of Theoretical Physics, \\
Joint Institute for Nuclear Research, Dubna 141980, Russia \\ [2mm]

$^2$Instituto de Fisica de S\~ao Carlos, Universidade de S\~ao Paulo, \\
CP 369, S\~ao Carlos 13560-970, S\~ao Paulo, Brazil} \\ [3mm]

{\bf E-mail}: yukalov@theor.jinr.ru 

\end{center}

\vskip 2cm

\begin{abstract}
Due to high current interest to Bose-Einstein condensation, the related topics, such as
spontaneous gauge symmetry breaking, ergodic decomposition, and particle fluctuations,
are intensively discussed in literature. These discussions, unfortunately, involve quite
a number of controversies and confusions. The goal of the present brief survey is to 
clarify some of these confusions, concentrating on the principal points, such as the
relationship between the conditions of Bose-Einstein condensation, the Bogolubov method 
of quasiaverages, spontaneous gauge symmetry breaking, ergodic decomposition, the presumed 
existence of nonthermodynamic particle fluctuations leading to the so-called ``grand 
canonical catastrophe", and the requirements for system stability.      

\end{abstract}

\vskip 2cm

\newpage

\section{Introduction}

The phenomenon of Bose-Einstein condensation (BEC) is currently attracting high attention 
with respect to both theory and experiment. Surprisingly, despite the long history of the 
topic, a number of points remains disputable and misunderstood. Among the often debated
problems, the following are discussed and confused over and over again. 

If there occurs BEC in the usual sense, without involving external fields, does this 
guarantee the existence of BEC in the sense of the Bogolubov quasiaverages?

If there appears BEC, whenever in the usual sense, or in the sense of the Bogolubov
quasiaverages, does this require global gauge symmetry breaking?

Does gauge symmetry breaking include the existence of BEC in the sense of the Bogolubov
quaisaverages? Are they equivalent? 
    
What is a correct form for the ergodic decomposition of a gauge invariant state into
states with broken gauge symmetry? 

Whether pathological nonthermodynamic condensate fluctuations, comprising the so-called 
``grand canonical catastrophe", can exist?  

The answers to these questions are highly important for the correct explanation of 
experiments that are so often misinterpreted. The aim of this brief survey is to present 
a concentrated explanation of the above mentioned points in a concise way. More details 
can be found in the cited literature. Throughout the paper, the grand canonical ensemble 
is employed.

\section{Space of states}

Let us consider a spinless Bose system characterized by the field operators $\psi({\bf r})$
and $\psi^\dagger({\bf r})$, obeying the Bose commutation relations and acting on a Fock 
space constructed as follows \cite{Berezin_1}. 

The vacuum vector is defined by the equation
\be
\label{1}
\psi(\br) \; |\; 0\; \rgl \; = \; 0 \;   .
\ee
Any vector of the space can be generated according to the action 
\be
\label{2}
|\; \vp \;\rgl \; = \;  
\sum_{n=0}^\infty \frac{1}{\sqrt{n!} } 
\int \vp(\br_1, \br_2,\ldots,\br_n) \;
\prod_{i=1}^n \psi^\dgr(\br_i)\; d\br_i \;|\; 0 \;\rgl \; .
\ee
The collection of such vectors composes a Fock space $\mathcal{F} = \mathcal{F}(\psi)$
generated by the field operator $\psi^\dagger({\bf r})$. A detailed exposition of the 
properties of the Fock space and akin topics is given in the review \cite{Yukalov_2}.

Field operators can be expanded over a basis $\{\varphi_k({\bf r})\}$,
\be
\label{3}
 \psi(\br)  \; = \;  \sum_k a_k \;\vp_k(\br) \; = \; \psi_0(\br) + \psi_1(\br) \;,
\ee
where the term $a_0 \varphi_0({\bf r})$, related to the possible condensate, is separated
from the terms related to noncondensed states,  
\be
\label{4}
  \psi_0(\br)  \; = \; a_0 \; \vp_0(\br) \; , 
\qquad
\psi_1(\br) \; = \; \sum_{k\neq 0} a_k \vp_k(\br) \;  .
\ee
The field operators $\psi_0$ and $\psi_1$ are orthogonal, such that
\be
\label{5}
\int \psi_0^\dgr(\br) \;\psi_1(\br) \; d\br \; = \; 0 \; ,
\ee
which follows from the orthogonality of the basis functions,
$$
\int \vp_k^*(\br) \;\vp_p(\br) \; d\br \; = \; \dlt_{kp} \; .
$$ 

The eigenvector of $\psi_0({\bf r})$,
\be
\label{6}
\psi_0(\br) \; |\; \eta \; \rgl \; = \; 
\eta(\br) \; |\; \eta \; \rgl \;  ,
\ee
is a normalized coherent state that is represented in the form
\be
\label{7}
|\; \eta \; \rgl \; = \; \exp\left\{ \; \int \left[\; 
 -\;\frac{1}{2} \;
|\;\eta(\br) \; |^2 + \eta(\br) \;\psi^\dgr(\br) 
\;\right] 
d\br \; \right\} \;| \; 0 \;\rgl \;  .
\ee

In what follows, let us deal with a uniform system of volume $V$, where
\be
\label{8}
 \psi_0(\br) \; = \; \frac{a_0}{\sqrt{V} } \; , \qquad
\eta(\br) \; = \; \frac{z}{\sqrt{V} } \; , \qquad 
a_0 \;|\; z \;\rgl \; = \; z \;|\;z \;\rgl \; ,
\ee
$z$ is a complex number related to the coherent state $|z \rangle$ which takes the form
\be
\label{9}
 |\;z \;\rgl \; = \exp\left( -\;\frac{1}{2} \; |\;z\;|^2 \right) 
\sum_{n=0}^\infty \frac{z^n}{\sqrt{n!} } \; | \; n \; \rgl \;  ,
\ee
with the $n$-particle vector
$$
| \; n \; \rgl \; = \; \frac{1}{\sqrt{n!}} \prod_{i=1}^n
\frac{1}{\sqrt{V}} \; \int \psi^\dgr(\br_i) \; d \br_i \;|\; 0 \;\rgl \;  .
$$
 
The set of all coherent vectors forms a Hilbert space $\mathcal{H}_0$, and its 
orthogonal compliment is denoted as $\mathcal{H}_1$. The Fock space $\mathcal{F}$ 
is isomorphic to the tensor product
\be
\label{10}
{\cal F} \; = \; \cH_0 \bigotimes \cH_1 \; . 
\ee

\section{Method of quasiaverages}

When the system grand Hamiltonian $H$ is invariant with respect to the global gauge 
symmetry, so that 
\be
\label{11}
 \hat U_\vp^+ \; H \; \hat U_\vp \; = \; H 
\qquad
\left( U_\vp = e^{i\vp\hat N} \right) \;  ,
\ee
with $\hat{N}$ being the number-of-particles operator, then the averaged field operator 
is zero,
\be
\label{12}
\lgl \;\psi(\br) \; \rgl \; = \; \
{\rm Tr}_{\cal F}\; \hat\rho \;\psi(\br) \; = \; 0 \; ,
\ee
due to the gauge invariance of the Gibbs statistical operator
\be
\label{13}
 \hat\rho \; = \;\frac{e^{-\bt H}}{{\rm Tr}_{\cal F} e^{-\bt H}} \; ,
\ee
in which $\beta$ is inverse temperature, $\beta = 1/T$. 

Here and in what follows the standard nonrelativistic Hamiltomian is assumed with a 
stable potential of pair interactions, such that the potential is a real even function 
of the difference of the positions of the two interacting particles. It may have a hard 
core, but needs to be locally square integrable on the complement of the closure of the 
hard core if there is any, and on the complement of the origin, if not.
 
In order to correctly describe the phenomena of phase transitions, such as Bose-Einstein 
condensation, Bogolubov \cite{Bogolubov_3,Bogolubov_4,Bogolubov_5,Bogolubov_6} developed 
the method of quasiaverages (see also \cite{Bogolubov_7}). The idea of the method is based 
on the understanding that the phase transition of BEC is accompanied by the gauge symmetry 
breaking. Hence, this symmetry needs to be broken in the theoretical description. For this 
purpose, the gauge symmetric Hamiltonian $H$ has to be complimented by a symmetry breaking 
term getting the symmetry broken Hamiltonian
\be
\label{14}
H_{\nu\al}\; = \; H + \nu \;\sqrt{V} \;\left( a_0^\dgr e^{i\al}
+ a_0 e^{-i\al} \right) \;   .
\ee
As is evident, the symmetry broken Hamiltonian reduces back to the symmetry invariant 
Hamiltonian, when $\nu$ tends to zero. The average of an operator $\hat{A}$ is calculated 
by averaging over the whole Fock space $\mathcal{F}$,
\be
\label{15}
 \lgl \;\hat A \;\rgl_{\nu\al} \;= \; 
{\rm Tr}_{\cal F}\; \hat\rho_{\nu\al} \;\hat A \; = \; 
{\rm Tr}_{\cH_0} {\rm Tr}_{\cH_1} \hat\rho_{\nu\al} \hat A \;  ,
\ee
where the symmetry broken statistical operator is
\be
\label{16}
\hat\rho_{\nu\al} \; = \; \frac{1}{Z_{\nu\al} } \; e^{-\bt H_{\nu\al} } \;,
\qquad
Z_{\nu\al} \; = \; {\rm Tr}_{\cal F}\; e^{-\bt H_{\nu\al} } \; .
\ee
The averaging over the space $\mathcal{H}_0$ reads as
$$
{\rm Tr}_{\cH_0} \hat\rho_{\nu\al} \hat A  \; = \;
\int \lgl \; z \;|\; \hat\rho_{\nu\al} \hat A \; | \; z \; \rgl \;
\frac{dz\; dz^*}{2\pi} \;  ,
$$
with the notation
$$
dz \; dz^* \; = \; 2d({\rm Re} z) \; d({\rm Im} z ) \;  ,
$$
so that the average (\ref{15}) is
\be
\label{17}
 \lgl \; \hat A \;\rgl_{\nu\al} \; = \; \int {\rm Tr}_{\cH_1}
\lgl \; z \;|\; \hat\rho_{\nu\al}\; \hat A \; |\; z \; \rgl \;
\frac{dz\; dz^*}{2\pi} \;  .
\ee

Using the relation
$$
z \; = \; |\; z \; | \; e^{i\vp} \; , \qquad
dz\; dz^* \; = \; 2 |\; z\;|\; d|\; z\;| \; d\vp \;   ,
$$
the average (\ref{17}) can be represented in the form
\be
\label{18}
\lgl \; \hat A \;\rgl_{\nu\al} \; = \; \int {\rm Tr}_{\cH_1}
 \lgl \; |\; z\;| e^{i\vp} \;|\; \hat\rho_{\nu\al}\; \hat A \; |\; |\; z \;| e^{i\vp} \; \rgl \;
2|\; z \;|\;d|\;z \;| \; \frac{d\vp}{2\pi} \;  .
\ee

If the system is finite and one sets $\nu = 0$, the average reduces back to the gauge 
symmetric invariant form with restored gauge symmetry,
\be 
\label{19}
\lim_{\nu\ra 0} \lgl \; \hat A \;\rgl_{\nu\al} \; = \;
{\rm Tr}_{\cal F}\;  \hat\rho \; \hat A \; \equiv \; \lgl \; \hat A \;\rgl \; .
\ee
It is, therefore, important to accomplish, first, the thermodynamic limit,
\be
\label{20}
N \; \ra \; \infty \; , \qquad
V \; \ra \; \infty \; , \qquad  
\frac{N}{V} \; \ra \; const \;  ,
\ee
obtaining the quasiaverage
\be
\label{21}
\lgl \; \hat A \;\rgl_\al \; = \; \lim_{\nu\ra 0} \; \lim_{V\ra\infty}
\lgl \; \hat A \;\rgl_{\nu\al} \;   ,
\ee
where the gauge symmetry remains broken after the limit.

It is assumed that in that limiting procedure the average of an operator $\hat{A}$ 
is finite. This requires that the operator be bounded, being from the class of local 
observables or appropriately normalized. For example, if the operator $\hat{A}$ 
corresponds to an extensive observable, it has to be taken in the form $\hat{A}/V$.   
  
It is important to emphasize that in the quasiaverages (\ref{15}) the averaging 
integration is always done over the variables of the whole space of states, with 
the differential measure $dz dz^*$.   

Note that, in principle, the symmetry breaking term can be removed not only after the 
thermodynamic limit, but simultaneously with, however a bit slower, than this limit, 
by using the Hamiltonian
\be
\label{22}
H_N \; =\; H + \frac{1}{N^\gm} \; \hat B \qquad ( 0 < \gm < 1 ) \; ,
\ee
where $\hat{B}$ is defined as an extensive observable breaking the gauge symmetry 
\cite{Yukalov_8}.

\section{Symmetry breaking}

The use of quasiaverages results in the broken gauge symmetry in the averages of operators 
as well as in thermodynamic quantities. The grand thermodynamic potential reads as
\be
\label{23}
\Om_{\nu\al} \; = \; - T \ln Z_{\nu\al} \; = \; - P_{\nu\al} V \; ,
\ee
with the pressure
\be
\label{24}
 P_{\nu\al} \; = \; \frac{T}{V} \; \ln Z_{\nu\al} \;  .
\ee
Since
$$
{\rm Tr}_{\cH_0} e^{-\bt H_{\nu\al} } \; = \;
\int \lgl\; z \;|\; e^{-\bt H_{\nu\al}} \;|\; z \;\rgl \;
\frac{dz \; dz^*}{2\pi} \;   ,
$$
the partition function reads as
\be
\label{25}
Z_{\nu\al} \; = \; 
\int  {\rm Tr}_{\cH_1} \lgl\; z \;|\; e^{-\bt H_{\nu\al}} \;|\; z \;\rgl \;
\frac{dz \; dz^*}{2\pi} \; .
\ee
Introducing the notation
\be
\label{26}
 e^{-\bt H_{\nu\al}(z)} \; = \; 
\lgl\; z \;|\; e^{-\bt H_{\nu\al}} \;|\; z \;\rgl
\ee
gives
\be 
\label{27}  
Z_{\nu\al} \; = \; 
\int  {\rm Tr}_{\cH_1 } e^{-\bt H_{\nu\al}(z)} \;
\frac{dz \; dz^*}{2\pi} \; .
\ee

Ginibre \cite{Ginibre_9} proved that, under the thermodynamic limit, when 
$V\ra \infty$,
\be
\label{28}
\int  {\rm Tr}_{\cH_1 } e^{-\bt H_{\nu\al}(z)} \;
\frac{dz \; dz^*}{2\pi} \; = \; 
\sup_z {\rm Tr}_{\cH_1 } e^{-\bt H_{\nu\al}(z)} \;  ,
\ee
so that the partition function becomes
\be
\label{29}
Z_{\nu\al} \; = \;  \sup_z {\rm Tr}_{\cH_1 } e^{-\bt H_{\nu\al}(z)} \;  .
\ee
The maximum of the partition function is reached at 
\be
\label{30}
 z_0 \; = \; \sqrt{N_0} \; e^{i\al} \; ,
\ee
when
\be 
\label{31}
\sup_z {\rm Tr}_{\cH_1 } e^{-\bt H_{\nu\al}(z)} \; = \;
{\rm Tr}_{\cH_1 } e^{-\bt H_{\nu\al}(z_0)} \; ,
\ee
$N_0$ being the number of particles in the condensate. The quasiaverage thermodynamic 
limit for pressure (\ref{24}), 
\be
\label{32}
\lim_{\nu\ra 0} \; \lim_{V\ra\infty} P_{\nu\al} \; = \; P_\al \; ,
\ee
is given by the expression 
\be
\label{33}
 P_\al \; = \; \lim_{\nu\ra 0} \; \lim_{V\ra\infty} \; \frac{T}{V} \;
\ln {\rm Tr}_{\cH_1 } e^{-\bt H_{\nu\al}(z_0)} \;  .
\ee
The necessary condition for the pressure maximum, or the thermodynamic potential 
minimum is
\be
\label{34}
 \frac{\prt P_\al}{\prt N_0} \; = \; 0 \;  ,
\ee
which can be written in the form
\be
\label{35}
\lim_{\nu\ra 0} \; \lim_{V\ra\infty} {\rm Tr}_{\cH_1 } \hat\rho_{\nu\al}(z_0) \;
 \frac{\prt H_{\nu\al}(z_0)}{\prt N_0} \; = \; 0 \;   ,
\ee
with the statistical operator
\be
\label{36}
 \hat\rho_{\nu\al}(z_0) \; = \; \frac{ \exp\{-\bt H_{\nu\al}(z_0)\} }
{ {\rm Tr}_{\cH_1 }\exp\{-\bt H_{\nu\al}(z_0) \}} \;  .
\ee

The use of quasiaverages breaks the gauge symmetry and explains the possibility of 
substitution, in the thermodynamic limit, of the operators $\psi_0$ and $a_0$ by the 
nonoperator quantities, according to the rule
\be
\label{37}
\psi_0 \;\ra \; \eta \; = \; 
\frac{z_0}{\sqrt{V}} \; = \; \sqrt{\rho_0} \; e^{i\al} \; ,
\qquad
a_0 \; \ra \; \sqrt{N_0} \; e^{i\al} \;  ,
\ee
as was suggested by Bogolubov \cite{Bogolubov_3,Bogolubov_4,Bogolubov_5,Bogolubov_6,Bogolubov_7}. 
The rigorous justification for this substitution was given by Ginibre \cite{Ginibre_9}. This
substitution can be done not necessarily for the ground-state, but for any mode and still 
gives the right answer even if the density $\rho_0 = N_0/V$ is small. However it becomes 
useful and principal only if $\rho_0$ is macroscopic and corresponds to nonzero condensate 
density.   

It is important to stress that the substitution (\ref{37}) breaks the gauge symmetry not 
merely by inducing the choice of a phase $\alpha$, but also it fixes the amplitude of the 
condensate variables, defining the condensate density as a maximizer of pressure or a minimizer 
of the thermodynamic potential. Spontaneous gauge symmetry breaking has sense because, by fixing 
the condensate phase and its amplitude, it makes the Bose-condensed system stable. Without
choosing the condensate amplitude as a minimizer of the thermodynamic potential, the system 
would be unstable.

\section{Bose-Einstein condensation}

A detailed and rigorous analysis of the relations between the conditions of BEC existence
and spontaneous gauge symmetry breaking has been done by Roepstorff \cite{Roepstorff_10},
Lieb, Seringer and Yngvasson \cite{Lieb_11,Lieb_12,Lieb_13}, and S\"{u}t\"{o} \cite{Suto_14}.
The studied relations concern the condition for the existence of BEC involving no external 
fields
\be
\label{38}
\lim_{V\ra\infty} \; \frac{\lgl\; a_0^\dgr a_0 \;\rgl}{V} \; > \; 0 \; ,
\ee
condition for the BEC existence in the sense of quasiaverages
\be
\label{39}
\lim_{\nu\ra 0} \; \lim_{V\ra\infty} \; 
\frac{\lgl\; a_0^\dgr a_0 \;\rgl_{\nu\al}}{V} \; > \; 0 \; ,
\ee
and the condition for the occurrence of spontaneous gauge symmetry breaking
\be
\label{40}
 \lim_{\nu\ra 0} \; \lim_{V\ra\infty} \; 
\frac{|\; \lgl\; a_0 \;\rgl_{\nu\al}\; |^2}{V} \; > \; 0 \;  .
\ee
   
It has been rigorously proved \cite{Lieb_11,Lieb_12,Lieb_13,Suto_14} that the description 
of BEC by means of the Bogolubov quasiaverages is equivalent to the occurrence of spontaneous 
gauge symmetry breaking,
\be
\label{41}  
\lim_{\nu\ra 0} \; \lim_{V\ra\infty} \; 
\frac{\lgl\; a_0^\dgr a_0 \;\rgl_{\nu\al}}{V} \; = \;
\lim_{\nu\ra 0} \; \lim_{V\ra\infty} \; 
\frac{|\; \lgl\; a_0 \;\rgl_{\nu\al}\; |^2}{V} \;   .
\ee
Moreover, it has been proved \cite{Roepstorff_10,Lieb_11,Lieb_12,Lieb_13} that the existence 
of BEC in the usual sense, that is without any external fields, means the existence of BEC 
in the sense of the Bogolubov quasiaverages,
\be
\label{42}
\lim_{V\ra\infty} \; \frac{\lgl\; a_0^\dgr a_0 \;\rgl}{V} \; \leq \;
\lim_{\nu\ra 0} \; \lim_{V\ra\infty} \; 
\frac{\lgl\; a_0^\dgr a_0 \;\rgl_{\nu\al}}{V} \;   .
\ee
The combination of equations (\ref{41}) and (\ref{42}) shows that the existence of BEC in
the usual sense implies the necessary occurrence of gauge symmetry breaking,
\be
\label{43}
\lim_{V\ra\infty} \; \frac{\lgl\; a_0^\dgr a_0 \;\rgl}{V} \; \leq \;
\lim_{\nu\ra 0} \; \lim_{V\ra\infty} \;
\frac{|\; \lgl\; a_0 \;\rgl_{\nu\al}\;|^2}{V} \;   .
\ee

In that way, the simultaneous appearance of BEC, involving quasiaverages and gauge symmetry 
breaking, is not an assumption but, on the contrary, the necessary consequence of the 
arising usual Bose condensate involving no external fields. In other words, the occurrence 
of the usual BEC, mentioning no quasiaverages, implies the existence of BEC in the sense 
of quasiaverages as well as the validity of the gauge symmetry breaking.

\section{Ergodic decomposition}

When the Hamiltonian and, consequently, the Gibbs state are invariant with respect to a 
group that can be decomposed into subgroups not allowing for their further decomposition,   
the averages of the operators of observables can be decomposed into a direct sum or integral
over the averages corresponding to the subgroups. This is called ergodic decomposition and 
it is correctly defined only in the thermodynamic limit
\cite{Ruelle_15,Dixmier_16,Emch_17,Bratteli_18}. In our case, the state, that is invariant
with respect to the gauge group $U(1)$, can be decomposed into the states with broken gauge
symmetry. 

In order not to confuse the ergodic decomposition with an incomplete integration, let us, 
first, explain why the latter has nothing to do with the broken-symmetry state. Let us for 
a moment forget quasiaverages and return back to the usual averages
\be
\label{44}  
 \lgl \; \hat A \;\rgl \; = \; 
{\rm Tr}_{\cH_0} \; {\rm Tr}_{\cH_1} \hat\rho \hat A  
\ee
defined in a finite volume $V$. The trace over $\mathcal{H}_0$ involves the integration 
over the complex numbers $z = |z| e^\varphi$. Denoting the incomplete integral 
\be
\label{45}
(\hat A)_\vp \; = \; \int {\rm Tr}_{\cH_1} \lgl\; 
|\; z\;| \; e^{i\vp} \; |\; \hat\rho \hat A \;| \; |\; z\;|\; e^{i\vp} 
\;\rgl\; 2|\; z\;| \; d|\;z\;| \; ,
\ee
containing the integration only over the modulus $|z|$, while no integration over the 
phase $\varphi$, it is straightforward to represent the average in the form
\be
\label{46}
 \lgl \; \hat A \;\rgl \; = \; \int_0^{2\pi} (\hat A)_\vp \;
\frac{d\vp}{2\pi} \;   .
\ee
This form, however, has nothing to do with the ergodic decomposition, which decomposes 
a symmetric average (state) into symmetry-broken averages (states). The incomplete
integral (\ref{45}) is not an average, it is merely a part of the whole average (\ref{44})
defined in the Fock space $\mathcal{F}(\psi)$, hence the incomplete integral (\ref{45})
is not a physically observable quantity. The equality (\ref{46}) is just an identity 
rewriting the definition of the average (\ref{44}) in a slightly different form. 

The correct ergodic decomposition requires to consider the thermodynamic limit. In this 
limit, as follows from the theorems \cite{Ginibre_9,Roepstorff_10,Lieb_11,Lieb_12,Lieb_13}, 
the Bogolubov representation, replacing the condensate field operator $\psi_0$ by the 
condensate function $\eta$, becomes asymptotically exact. This implies that the operator 
sum (\ref{4}) takes the form 
\be
\label{47}
\psi(\br) \; = \; \eta(\br) + \psi_1(\br)   
\ee
called the Bogolubov shift. The condensate function and the field operator of non-condensed
particles are mutually orthogonal,
\be
\label{48}
\int \eta^*(\br) \; \psi_1(\br) \; d\br \; = \; 0 \; .
\ee
The condensate function is prescribed to be the minimizer of the thermodynamic potential 
or the maximizer of the pressure. 

In the thermodynamic limit, the system Fock space is not the same space (\ref{10}) as that 
for a finite volume. Introducing the vacuum state by the condition
\be
\label{49}
 \psi_1(\br) \;|\; 0 \;\rgl_1 \; = \; 0 \;  ,
\ee
it is straightforward to create the vectors
\be
\label{50}
 |\; \vp_1 \;\rgl \; = \; \sum_{n=0}^\infty 
\frac{1}{\sqrt{n!}} \int \vp_1(\br_1,\br_2,\ldots,\br_n) \;
\prod_{i=1}^n \psi_1(\br_i) \; d\br_i \;|\; 0 \;\rgl_1 \; .
\ee
The family of all these states forms the Fock space $\mathcal{F}(\psi_1)$, generated by the
field operator $\psi_1^\dagger({\bf r})$. The properties of this Fock space are described, 
e.g., in the articles \cite{Yukalov_19,Yukalov_20,Yukalov_21}. The vacuum $|0\rangle_1$ is 
a coherent state $|\eta>$, or, more correctly, a set of the coherent states 
\be
\label{51}
|\; \eta_\al \;\rgl \; = \; |\; |\;\eta \;| \; e^{i\al} \; \rgl 
\qquad ( 0 \leq \al\leq 2\pi)
\ee
differing by the phases. Respectively, the field operators of non-condensed particles can 
generate from the family of the vacua, represented by the coherent states with different 
phases, a family of Fock spaces, which are orthogonal to each other \cite{Yukalov_21}. 
In other words, the Fock space generated by the operator $\psi$ at finite volume decomposes,
in the thermodynamic limit, into a collection of spaces $\mathcal{F}_\alpha$ generated from 
the vacua with different phases,
\be
\label{52}
 {\cal F}(\psi) \; \mapsto \; \int^\oplus {\cal F}_\al \; \frac{d\al}{2\pi} \;  ,
\ee
as $V \ra \infty$. The spaces $\mathcal{F}_\alpha$ with different values of the phase 
$\alpha$ are mutually orthogonal, and each space $\mathcal{F}_\alpha$ is orthogonal to 
the space $\mathcal{F}(\psi)$.  

In each space $\mathcal{F}_\alpha$, it is possible to define the quasiaverages 
$\langle \hat{A} \rangle_\alpha$ with broken gauge symmetry, characterized by different 
fixed phases. This gives the ergodic decomposition of the invariant averages
\be
\label{53}
 \lgl \; \hat A \;\rgl_{{\cal F}(\psi)} \; \mapsto \; 
\int  \lgl \; \hat A \;\rgl_\al \; \frac{d\al}{2\pi} \; , 
\ee
as $V \ra \infty$, where the left-hand side is invariant over gauge transformations, while 
in each term of the right-hand-side integral the gauge symmetry is broken. Thus an average, 
invariant with respect to a group of gauge transformations, can be decomposed into an integral 
of the ergodic terms with broken gauge symmetry, where the integrands are given by the
Bogolubov quasiaverages \cite{Yukalov_22,Yukalov_23,Verbeure_23,Wreszinski_23}.   

It is important to stress that the separation of a state with a fixed phase $\alpha$ is
caused by the action of the infinitesimal source breaking the gauge symmetry, but not to
the incomplete integration omitting the integration over the phase $\varphi$. The 
definition of a quasiaverage includes the integration over all variables characterizing 
the state, including the phase $\varphi$. The reduction of the variables, when passing 
from the space $\mathcal{F} = \mathcal{H}_0 \bigotimes \mathcal{H}_1$ to the variables
related only to the space $\mathcal{H}_1$ is due to the property (\ref{28}) imposing
the condition of choosing the condensate variable $z$ so that to maximize the partition
function (\ref{29}).   
 
Together with the phase transition of Bose condensation, there happens the change in 
the structure of the phase space characterizing the physical system. A normal system in
a finite volume is described in the space (\ref{10}). Under the increasing volume and 
appearing BEC, the space transforms into the space (\ref{52}). Commutation relations of
the field operators $\psi({\bf r})$ and $\psi_1({\bf r})$ are the same Bose commutation 
relations, but these operators act in different spaces. This is what is called unitary 
inequivalent representations of commutation relations \cite{Araki_23} (see also 
\cite{Yukalov_19,Yukalov_20,Yukalov_21}). Physically, the choice of one of the spaces 
$\mathcal{F}_\alpha$ occurs as a result of a random fluctuation. Mathematically, this is 
due to the spontaneous symmetry breaking realized through quasiaverages. Thus the change 
of the physical system, under the appearance of BEC, is accompanied by the following space 
transformations: 
\be
\label{54}
 {\cal F}(\psi) \; = \; 
\cH_0 \bigotimes \cH_1 \; \mapsto \; 
\int^\oplus {\cal F}_\al \; \frac{d\al}{2\pi} \; .
\ee

\section{Condensate fluctuations}

In order that the theoretical description of the arising BEC would be adequate to real 
physical processes, it is very important to understand what is the space of states of 
the system and accomplish calculations in that space. Without a clear understanding of
the frames prescribed by the theory, one can come to meaningless conclusions. A sharp 
example of such a confusion is the so-called ``grand canonical catastrophe", which is
just a result of forgetting to break the gauge symmetry. Recall that the spontaneous 
gauge symmetry breaking and the related Bogolubov quasiaverages necessarily arise as soon 
as Bose-Einstein condensation occurs \cite{Bogolubov_3,Bogolubov_4,Bogolubov_5,Bogolubov_6,
Bogolubov_7,Ginibre_9,Roepstorff_10,Lieb_11,Lieb_12,Lieb_13,Suto_14}. This is not an 
assumption but a conclusion based on rigorous mathematical theorems.

The so-called ``grand canonical catastrophe" appears in studying condensate fluctuations 
in a Bose-condensed system, but without gauge symmetry breaking \cite{Fujiwara_24,Ziff_25},
when the variance of the condensate number of particles is proportional to $N^2$. Such 
pathologically catastrophic fluctuations are, of course, unphysical, being merely caused 
by considering Bose-Einstein condensation and forgetting to break the gauge symmetry
\cite{Ter_26}. A nice overview of the history and a detailed description of conflicting
opinions is given by Kruk et al. \cite{Kruk_27}.

When the gauge symmetry is broken, which is compulsory under the existence of BEC, there 
are no pathological fluctuations, as has been demonstrated in the reviews 
\cite{Yukalov_28,Yukalov_29}. Moreover, employing the Bogolubov quasiaverages results in 
negligible condensate fluctuations. This can be clearly illustrated for the ideal Bose gas, 
with the grand Hamiltonian
\be
\label{55}
H \; = \; \sum_k \om_k \; a_k^\dgr \; a_k 
\qquad
\left( \om_k =\frac{k^2}{2m} - \mu \right) \;   .
\ee
Adding the terms, breaking the Hamiltonian gauge symmetry, gives
\be
\label{56}
H_{\nu\al} \; = \; \sum_k \om_k \; a_k^\dgr \; a_k  +\nu\; \sqrt{V}\;
\left( a_0^\dgr e^{i\al} + a_0 e^{-i\al} \right) \; .
\ee
The number of condensed particles is
\be
\label{57}
 N_0 \; = \; \lgl\; \hat N_0 \; \rgl_{\nu\al} \qquad
(\hat N_0 = a_0^\dgr a_0 ) \;  .
\ee
And the variance of the number of condensed particles is
\be
\label{58}
{\rm var}(\hat N_0) \; = \; \lgl\; \hat N_0^2 \; \rgl_{\nu\al} -
 \lgl\; \hat N_0 \; \rgl_{\nu\al}^2 \; = \; 
T \; \frac{\prt N_0}{\prt\mu} \;  .
\ee
For the number of condensed particles, taking into account that $\mu$ is small, one has 
\be
\label{59}
 N_0 \; = \; -\; \frac{T}{\mu} + \frac{\nu^2}{\mu^2} \; V \;  ,
\ee
while the variance reads as
\be
\label{60}
{\rm var}(\hat N_0) \; = \; \frac{T^2}{\mu^2}\; - \; 
\frac{2T\nu^2}{\mu^3} \; V \; .
\ee
According to the definition of quasiaverages, one has, first, to take the thermodynamic 
limit for the observable quantity
\be
\label{61}
 \lim_{V\ra\infty} \; \frac{{\rm var}(\hat N_0)}{V} \; = \; - \; 
\frac{2T}{\mu^3} \; \nu^2 \;   ,
\ee
after which to take the limit $\nu \ra 0$, obtaining
\be
\label{62}
\lim_{\nu\ra 0} \; \lim_{V\ra\infty} \; 
\frac{{\rm var}(\hat N_0)}{V} \; = \; 0 \;  .
\ee

This conclusion can be achieved even simpler, remembering the mathematical theorems
\cite{Ginibre_9,Roepstorff_10,Lieb_11,Lieb_12,Lieb_13,Suto_14} stating that, in the 
thermodynamic limit, the Bogolubov replacement of the condensate operator $a_0$ by a
nonoperator term $\sqrt{N_0} e^{i \alpha}$ becomes exact. Therefore
\be
\label{63}
\lim_{\nu\ra 0} \; \lim_{V\ra\infty} \;  
\frac{{\rm var}(a_0^\dgr a_0)}{V} \; = \; 
\lim_{\nu\ra 0} \; \lim_{V\ra\infty} \; 
\frac{{\rm var}(N_0)}{V} \; = \; 0 \; .
\ee

Thus, when correctly describing the condensate fluctuations, not forgetting the gauge 
symmetry breaking, there are neither pathological condensate fluctuations, nor any 
``grand canonical catastrophe".

\section{Stability conditions}

It is quite surprising that those who talk about catastrophic condensate fluctuations
under the so-called ``grand canonical catastrophe" keep silence with regard to stability
conditions of statistical systems. In order that an equilibrium statistical system be 
stable, its susceptibilities with respect to the variation of thermodynamic quantities 
have to be positive and finite, but negative or infinite susceptibilities signify 
instability \cite{Landsberg_30}. The susceptibilities are expressed through the variances 
of the operators of observable quantities describing the fluctuations of the related 
observables. Let a self-adjoint operator $\hat{A}$ represent an observable quantity. The 
fluctuations of the observable are characterized by the variance
\be
\label{64}
 {\rm var}(\hat A) \; = \; 
\lgl\; ( \hat A - \lgl\; \hat A \; \rgl)^2 \; \rgl \; = \; 
\lgl\;  \hat A^2 \;\rgl - \lgl\;  \hat A \;\rgl^2 \; .
\ee 
Fluctuations in an equilibrium system concern a stable system provided they are 
thermodynamically normal, implying that
\be
\label{65}
 0\; \leq \; \frac{ {\rm var}(\hat A)}{N} \; < \; \infty \;  .
\ee
This means that in the thermodynamic limit, where the fluctuations of an observable 
quantity behave as
\be
\label{66}
{\rm var}(\hat A) \; \simeq \; c N^\zeta 
\qquad ( N \ra \infty) \; ,
\ee
with $c$ being a constant independent on $N$, the fluctuation index \cite{Yukalov_31} 
is such that $\zeta \leq 1$. 

If the observable quantity corresponds to a composite operator
\be
\label{67}
\hat A \; = \; \sum_i \hat A_i \;   ,
\ee
with the variances of the partial operators
\be
\label{68}
{\rm var}(\hat A_i ) \; \simeq \; c_i N^{\zeta_i} 
\qquad ( N \ra \infty) \;  ,
\ee
then the fluctuation theorem \cite{Yukalov_32,Yukalov_33} states that
\be
\label{69}
\zeta  \; = \; \sup_i \zeta_i \;   .
\ee
This tells us that the fluctuations of a composite operator are thermodynamically normal 
if and only if the fluctuations of all partial operators are normal, and they are
thermodynamically anomalous, meaning the system instability, when at least one of the 
partial variances is thermodynamically anomalous.  

For a system with BEC, the fluctuations of density are the most interesting. The relative
density fluctuations, measured by the ratio $\rm{var}(\hat{N})/N$, are directly observable,
entering such observable quantities as the isothermal compressibility
\be
\label{70}
 \varkappa_T \; = \; \left( \frac{1}{\rho T}\right) \;
\frac{{\rm var}(\hat N)}{N} \;  ,
\ee
long-wave structure factor  
\be
\label{71}
\lim_{k\ra 0} S(k) \; = \;  \frac{{\rm var}(\hat N)}{N} \; ,
\ee
and sound velocity squared
\be
\label{72}
 s^2  \; = \; 
\frac{T}{m} \; \left[\; \frac{{\rm var}(\hat N)}{N} \;\right]^{-1} \; .
\ee
Note that the above definitions are the exact thermodynamic relations. 

If the ``grand canonical catastrophe" would exist, the condensate fluctuations would be 
catastrophic, being proportional to $N^2$. According to the fluctuation theorem 
\cite{Yukalov_32,Yukalov_33}, the operator of the total number of particles 
$\hat{N} = \hat{N}_0 + N_1$ would also exhibit the variance proportional to $N^2$. 
This means that for a large number of particles $N \gg 1$ the relative variance 
${\rm var}(\hat{N})/N$ would diverge as $N \ra \infty$. This would imply that the
compressibility diverges, the structure factor diverges, and the sound velocity is zero. 
It is necessary to stress that this divergence occurs independently of thermodynamic 
parameters, diverging with respect to $N \ra \infty$. It is clear that such a system 
cannot be stable. Then neither dilute Bose-condensed gases nor superfluid liquids could 
exist in a stable equilibrium state. As far as both, quantum Bose gases as well as 
superfluid liquids, such as Helium-$4$, do exist in equilibrium state, there follows 
the sole conclusion that the so-called ``grand canonical catastrophe" does not exist, 
which is in agreement with the theory of BEC with broken gauge symmetry.

\section{Conclusion}

The theory of BEC, based on obvious mathematical facts, leads to the following 
conclusions: 
 
\begin{itemize}

\item
The Bogolubov substitution (\ref{37}) of the condensate field operator by a nonoperator 
term, defined as a minimizer of the thermodynamic potential, is asymptotically exact in 
the thermodynamic limit. 

\item
If there occurs BEC without external fields, then there occurs BEC characterized by the
Bogolubov quasiaverages (\ref{42}).

\item
The existence of BEC in the sense of the Bogolubov quasiaverages is equivalent to the gauge
symmetry breaking (\ref{41}). 

\item 
The existence of BEC without external fields implies the gauge symmetry breaking in terms 
of quasiaverages. The main relations between the condensate existence and gauge symmetry 
breaking is shown in Figure 1.

\item
The incomplete integration (\ref{45}) does not define an average, and equality (\ref{46}) 
is not an ergodic decomposition, but just the same identity as (\ref{44}) defining the
gauge invariant average in a finite volume. The ergodic decomposition has sense only in 
the thermodynamic limit and with the averages containing the integration over the 
corresponding whole spaces of states. 

\item
The ergodic decomposition (\ref{53}), which is meaningful only in the thermodynamic limit, 
decomposes the gauge invariant state over the set of pure states with broken gauge symmetry. 
Each state in the integral (\ref{53}) is defined in the mutually orthogonal spaces of states 
with different phases of the condensate. The integrand averages of this decomposition are 
the Bogolubov quasiaverages. 

\item
Condensate fluctuations, under the broken gauge symmetry, are negligible, as is proved in 
equations (\ref{62}) and (\ref{63}). 

\item
The so-called ``grand canonical catastrophe" does not exist, being just a result of incorrect 
calculations assuming the occurrence of BEC, but forgetting that the existence of BEC, even
without mentioning quasiaverages, necessarily breaks the gauge symmetry and affirms the
validity of the Bogolubov quasiaverages.  

\item   
The occurrence of the ``grand canonical catastrophe" is impossible, since it would mean the 
system instability, with a divergent compressibility (\ref{70}), divergent structure factor 
(\ref{71}), and zero sound velocity in (\ref{72}).

\item
Equilibrium statistical systems are stable if their fluctuations are thermodynamically 
normal in the sense of (\ref{65}). 
\end{itemize}

If, according to rigorous theory, nonthermodynamic, or even catastrophic, condensate 
fluctuations cannot exist in stable equilibrium systems, then there arises a question: 
How to understand the claims that in some experiments one has observed such 
nonthermodynamic fluctuations? As far as what is strictly proved in the theory to be 
nonexistent cannot be observed, it is necessary to look for other reasons. For instance,
in experiments there are finite-size effects, boundary effects, initial-condition 
effects, and various technical noise. It is also very likely that the studied system 
is not in a state of stable equilibrium. However, the analysis of particular details 
of separate experiments is out of the scope of the present paper. The most important is 
to keep in mind that what is mathematically proved to occur, certainly does occur, and 
what is proved to be absent, for sure does not exist.   

%Figure 1
\begin{figure}[ht]
\centerline{
\includegraphics[width=10cm]{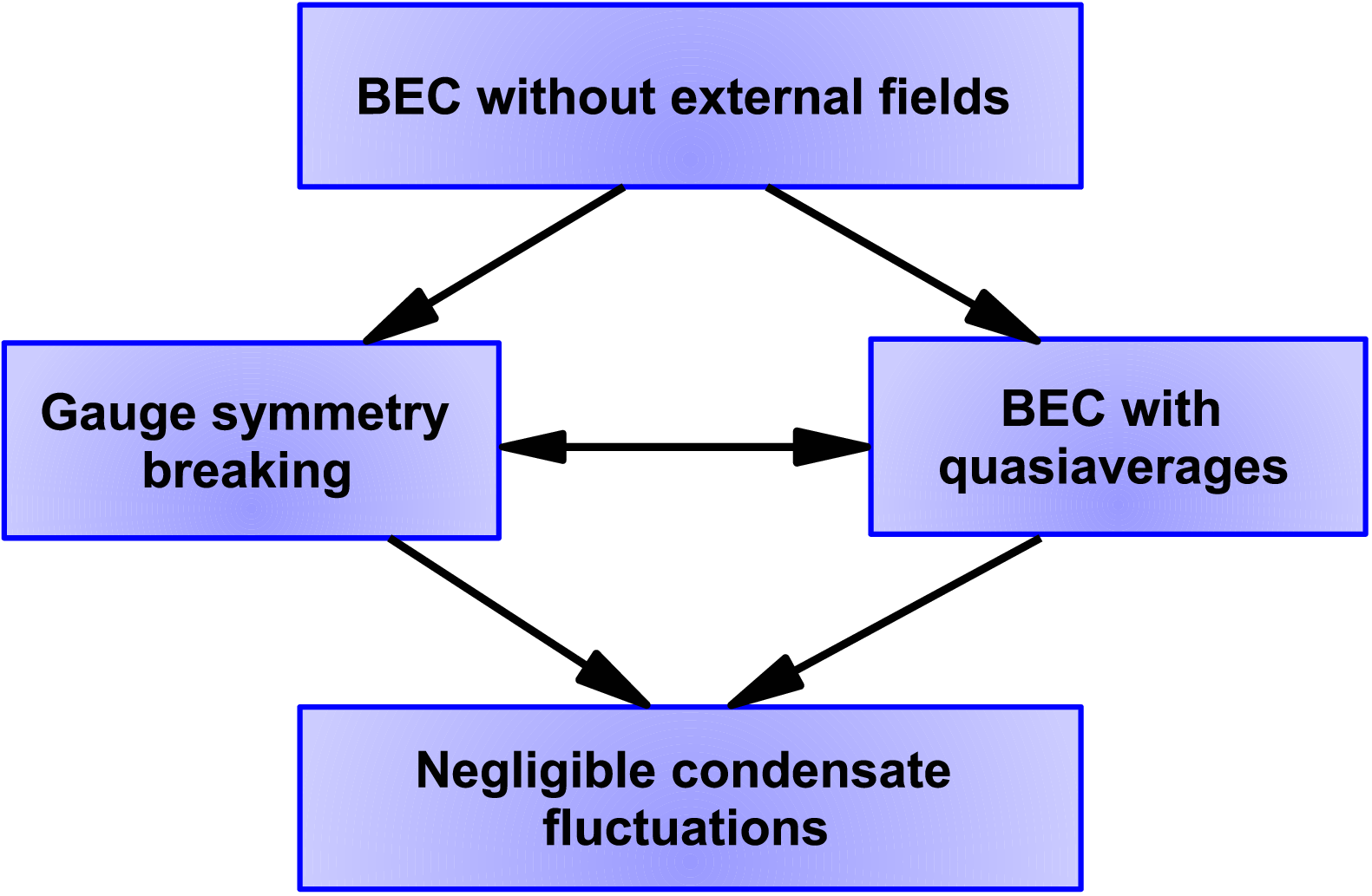} }
\caption{\small
The relations between the BEC occuring without external fields,
BEC in the sense of the Bogolubov quasiaverages, gauge symmetry breaking,
and condensate fluctuations.
}
\label{fig:Fig.1}
\end{figure} 

\vskip 5mm    
 
\section*{Data availability statement}

No new data were created or analyzed in this study. Data sharing is not applicable to 
this article.

\section*{Funding}

This research received no external funding

\section*{Conflicts of interest}

The authors declare no conflict of interest.

\section*{Author contribution}

Conceptualization, V Y; Methodology, V Y; Validation, V Y; Formal Analysis, V Y; 
Investigation, V Y; Writing--Original Draft Preparation, V Y; Writing--Review $\&$ 
Editing, V Y; Supervision, V Y.

\end{document}